\documentclass{appolb}
\usepackage{amsmath,amssymb,graphicx,xcolor}

\newcommand{\order}{\mathcal{O}}
\renewcommand{\Re}{\operatorname{Re}}

\begin{document}
\title{Heavy quark thermodynamics with anisotropic lattices%
\thanks{Presented at Excited QCD, Granada, 8--14 January 2026}%
}
\author{Jon-Ivar Skullerud, Rachel Horohan D'Arcy
\address{Department of Physics, National University of Ireland
  Maynooth,\\ Maynooth, County Kildare, Ireland}\\[3mm]
Gert Aarts, Chris Allton, M. Naeem Anwar, Timothy J. Burns, Ben Page
\address{Centre for Quantum Fields and Gravity, Department of Physics,\\ Swansea University, Swansea, SA2 8PP, United Kingdom}\\[3mm]
Ryan Bignell, Sin\'ead M. Ryan
\address{School of Mathematics and Hamilton Mathematics Institute,\\
  Trinity College Dublin, Dublin, Ireland}\\[3mm]
Benjamin J\"ager
\address{Quantum Field Theory Center \& Danish IAS, Department of Mathematics and Computer Science,
  University of Southern Denmark, 5230 Odense M, Denmark}\\[3mm]
Seyong Kim
\address{Department of Physics, Sejong University, Seoul 05006, Korea}\\[3mm]
Maria Paola Lombardo
\address{INFN, Sezione di Firenze, 50019 Sesto Fiorentino (FI), Italy}\\[3mm]
Alexander Rothkopf
\address{Department of Physics, Korea University, Seoul 02841, Korea}\\[3mm]
Antonio Smecca
\address{INFN,
  Sezione di Roma Tre, Via della Vasca Navale 84, I-00146 Rome,
  Italy\\
Centre for Quantum Fields and Gravity, Department of Physics,\\ Swansea University, Swansea, SA2 8PP, United Kingdom}
}
\maketitle
\begin{abstract}
We present recent results from the \textsc{Fastsum} collaboration, using
anisotropic lattice QCD to study spectral properties of heavy
quarkonia and open heavy flavour systems at high temperature. For
heavy quarkonium, our results using a number of different methods suggest a small but significant and robust negative mass shift as well as an increasing thermal width. We present the first lattice results for masses and spectral functions of $B$ mesons at high temperature, and preliminary results for a high-precision calculation of the static quark potential.
\end{abstract}
  
\section{Introduction}

Heavy quarks are among the most prominent probes of heavy-ion
collisions \cite{Andronic:2024oxz}: due to their large mass they are
predominantly created in the initial hard collisions and subsequently
experience the full space–time evolution of the plasma. Their
interactions with the hot medium encode key information about
transport properties such as diffusion coefficients and energy loss
mechanisms. Furthermore, the partial thermalization of heavy quarks and
their participation in collective flow provide insight into the degree
of coupling within the QGP. 

The properties of the quarks and their bound states in the medium are
encoded in spectral functions which are related to the euclidean
correlators that can be computed on the lattice by a (generalised)
Laplace transform.  Inverting this relation to obtain the spectral
function given the euclidean correlator is a well-known ill-posed problem.
This problem may be ameliorated by using anisotropic lattices with a
finer resolution in (imaginary) time than in the spatial directions.
The \textsc{Fastsum} collaboration has pioneered the use of
anisotropic lattices for this purpose.  In these proceedings we will
present recent results on three topics related to heavy quarks in the
medium: thermal mass shift and width of bottomonium, open-beauty
mesons, and the real-time static quark potential.

\section{Lattice setup}
\label{sec:lattice}

Our simulations are carried out using anisotropic lattices with an
$\order(a^2)$ improved gauge action and an $\order(a)$ improved Wilson
fermion action with stout smearing, following the parameter tuning and ensembles generated by the Hadron Spectrum Collaboration \cite{Edwards:2008ja,Lin:2008pr}.  The results presented here were
produced using the ``Gen2'' and ``Gen2L'' ensembles, which have $N_f=2+1$ active
quark flavours with $m_\pi\approx380$ and 240\,MeV respectively and an approximately
physical strange quark.  The spatial lattice spacing is
$a_s=0.1205(8)\,$fm (Gen2) and $0.1121(3)\,$fm (Gen2L), and the anisotropy
$\xi=a_s/a_\tau=3.45$.  The 
temperature is given by $T=(a_\tau N_\tau)^{-1}$ and is varied by
changing the number of sites $N_\tau$ in the temporal direction.  For
more details about the ensembles, see
Ref.~\cite{Aarts:2014nba,Aarts:2020vyb,Aarts:2022krz} and references therein.

\begin{table}[h]
\small
\setlength{\tabcolsep}{4pt} 
\begin{tabular}{cc|ccccccccccc}
& $N_\tau$ & 128 & 64 & 56 & 48 & 40 & 36 & 32 & 28 & 24 & 20 &
  16\\\hline
  Gen2 & $T$ (MeV) & 44 & & & 117 & 141 & 156 & 176 & 201 & 235 & 281
  & 352 \\
  & $T/T_c$ & 0.24 & & & 0.65 & 0.78 & 0.86 & 0.97 & 1.11 & 1.3 & 1.6 & 1.9 \\\hline
  Gen2L & $T$ (MeV) & 47 & 95 & 109 & 127 & 152 & 169 & 190 & 217 &
  253 & 304 & 380 \\
  & $T/T_c$ & 0.28 & 0.57 & 0.65 & 0.76 & 0.91 & 1.01 & 1.14 & 1.3 &
  1.5 & 1.8 & 2.3\\\hline
\end{tabular}
\caption{Lattice temporal extents $N_\tau$ and temperatures $T$ used
  in this study, in units of MeV and of the chiral transition
  temperature $T_c$.}
\label{tab:temperatures}
\end{table}

\section{Thermal mass and width of quarkonium}
\label{sec:quarkonium}

Using our Gen2L ensembles, we have compared a number of different
methods for determining the thermal mass shift and width of heavy
quarkonia.  We have simulated the $b$ quarks using an NRQCD action
including $\order(v^4)$ and leading spin-dependent corrections, with
mean-field improved tree-level coefficients \cite{Aarts:2014cda}.  We
have employed direct correlator analysis using time-derivative moments
\cite{Darcy:2025tzj} and a generalised eigenvalue problem (GEVP)
analysis \cite{Bignell:2025bga}, linear methods: Tikhonov,
Backus--Gilbert and Hansen--Lupo--Tantalo (HLT) \cite{Smecca:2025hfw},
and bayesian methods: maximum entropy (MEM) and the BR method
\cite{Burnier:2013nla}.  Further details are given in
\cite{Skullerud:2025iqt}.

\begin{figure}
\includegraphics[width=0.49\textwidth]{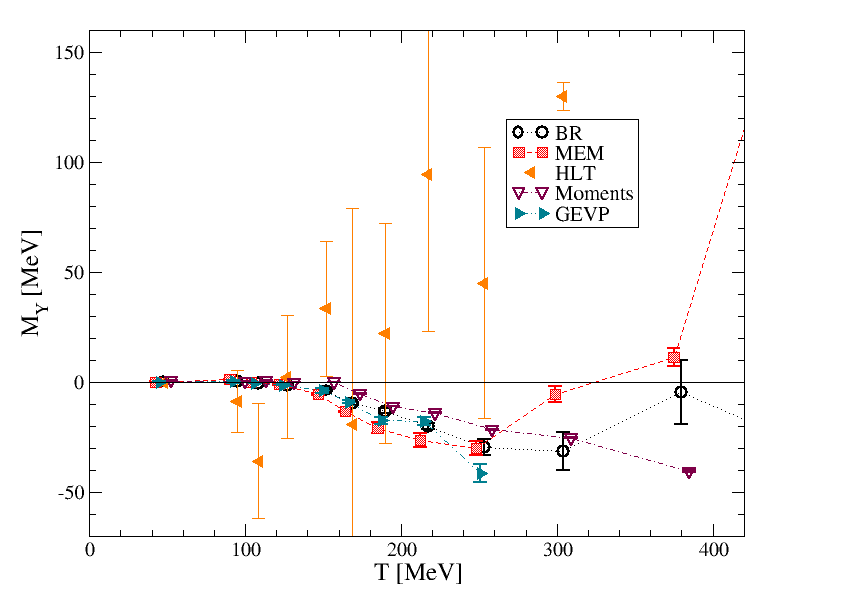}    
\includegraphics[width=0.49\textwidth]{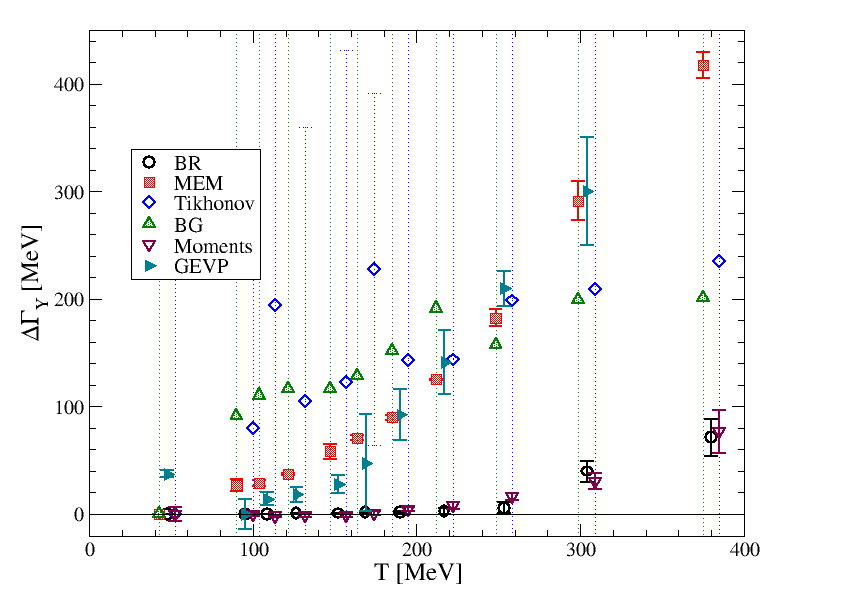} 
\caption{The thermal mass shift (left) and width (right) of the
  $\Upsilon(1S)$ using different methods.  The results show the
  difference between the thermal and equivalent zero-temperature
  results, see text for details.}
\label{fig:Upsilon}
\end{figure}

Figure~\ref{fig:Upsilon} shows the thermal mass shift and width of the
$\Upsilon(1S)$ from the different methods.  Where possible, these have
been determined by applying the same analysis to the correlators from
our coldest lattice, $N_\tau=128$ (which we term $T=0$), with temporal
extent truncated to match that of the corresponding thermal
correlator.  The resulting mass and width have been subtracted from
the thermal result to give the mass shifts and widths shown.  Where
this analysis has not yet been performed, the values obtained from the
$T=0$ ensemble with the full temporal extent have been subtracted.

We see that the linear methods have much larger uncertainties than
the others, which is not surprising as they are not designed to
identify narrow peaks in spectral functions (although they become
exact in the continuous-time limit).  The other methods all
find a small but significant negative mass shift of up to 40 MeV.  All
methods show a thermal width that increases with temperature, but they
differ on its magnitude, so at present we can only place an upper
bound on the width.

\section{B mesons}
\label{sec:Bmesons}

\begin{figure}
\includegraphics[width=0.49\textwidth,height=4.2cm]{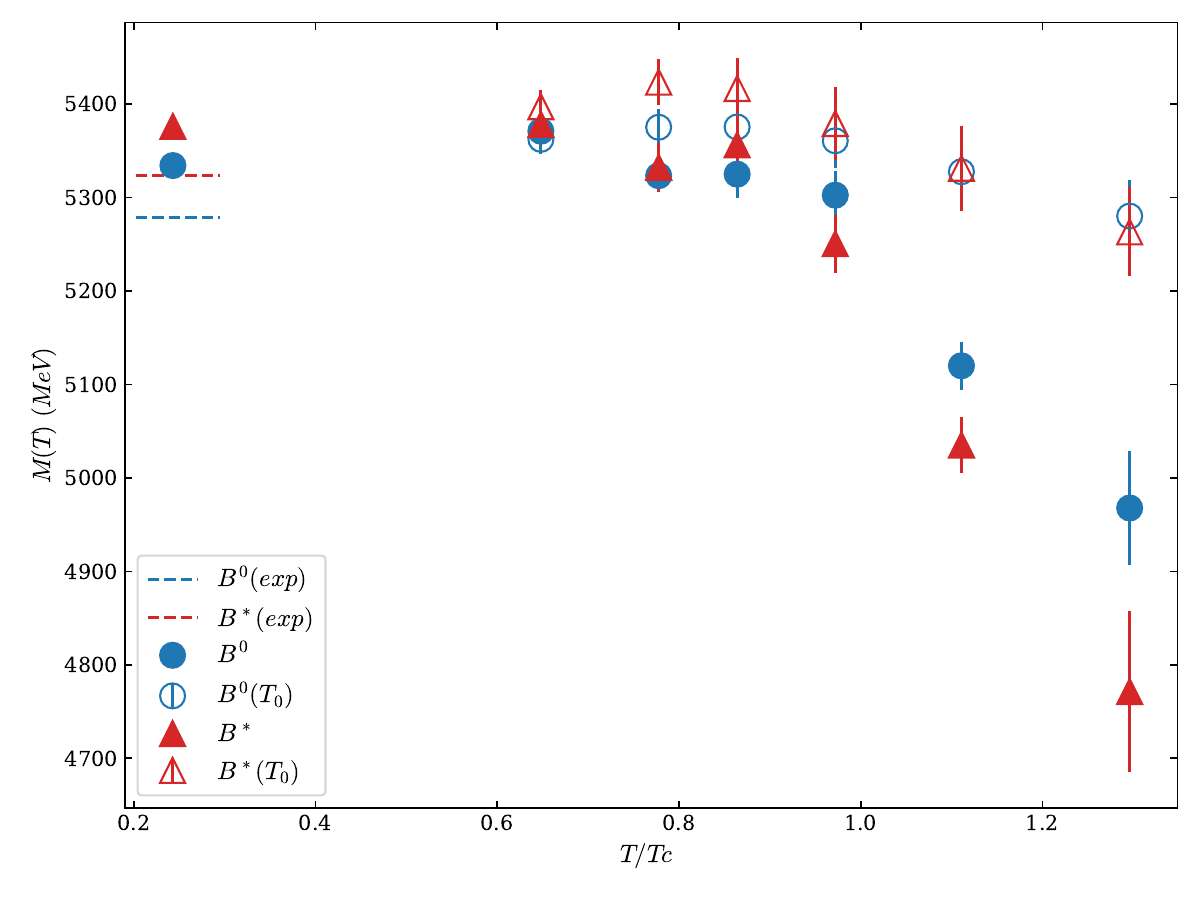}
\includegraphics[width=0.49\textwidth,height=4.2cm]{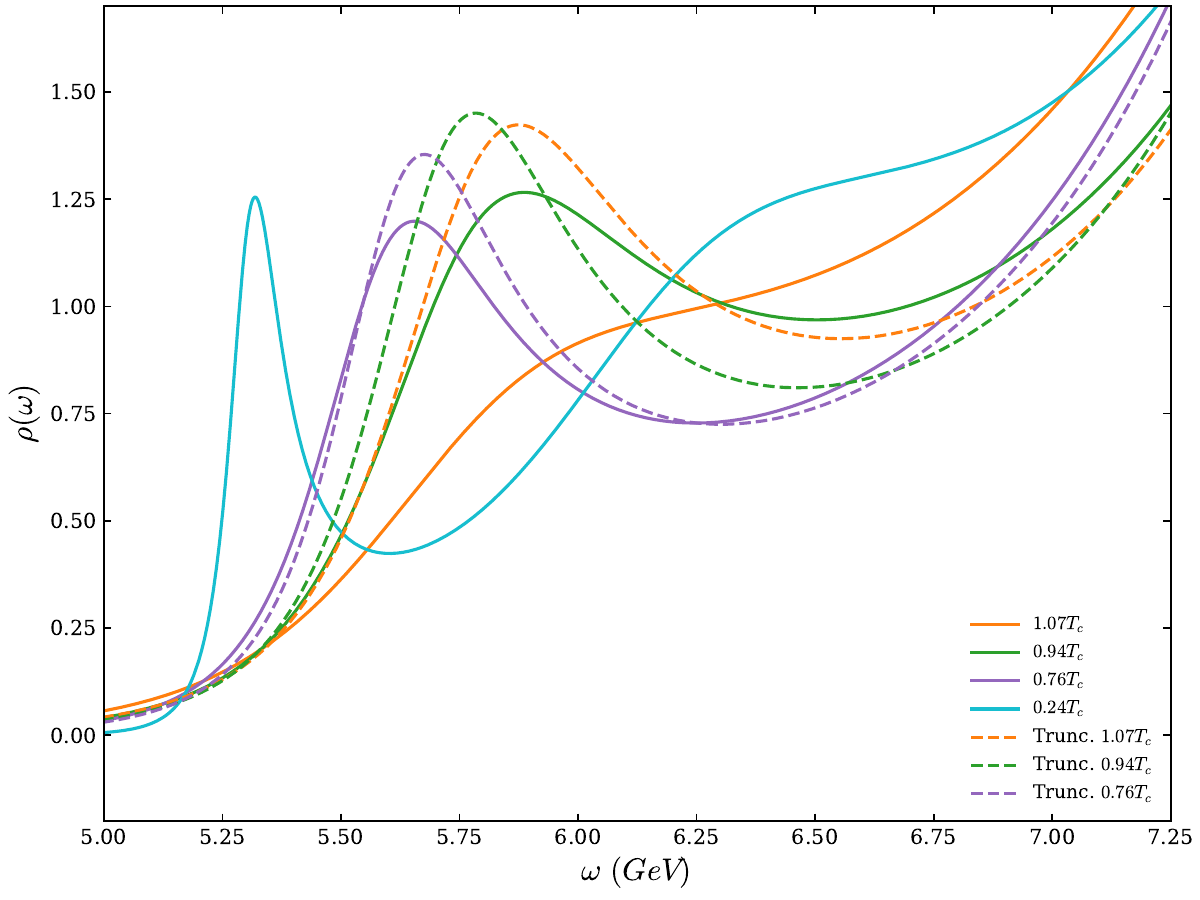}
\caption{Left: The mass of the $B$ and $B^*$ mesons as a function of
  temperature, compared to the masses extracted from $T=0$ correlators
truncated to the same temporal extent.  Right: $B$ meson spectral
functions from the BR method (solid lines), together with spectral functions
obtained from truncated $T=0$ correlators (dashed lines).}
\label{fig:Bmesons}
\end{figure}

We constructed two-point correlation functions for B-mesons by
combining NRQCD propagators for the $b$ quark with relativistic
propagators for the light anti-quark.  The two are combined by tracing
the upper two Dirac indices (in the nonrelativistic representation) of
the light (anti-)quark propagator with the full NRQCD propagator and
the appropriate spin matrices to produce pseudoscalar and vector states.


The light quark propagator is computed with antiperiodic boundary
conditions in the temporal extent while the NRQCD heavy quark
propagator satisfies an initial condition problem. This results in the
light quark propagator containing both forward and backward
propagation and the NRQCD heavy quark propagator only forward
propagation. To ensure we only considering the forward propagating
light antiquark we restrict our analysis to $\tau\leq N_\tau/2$.

Figure~\ref{fig:Bmesons} (left) shows the mass of the $B$
(pseudoscalar) and $B^*$ (vector) mesons determined from standard
exponential fits, as function of temperature, from the Gen2 ensembles.
The masses are higher than the experimental values since we are using
heavier-than-physical light quarks.
We also show the results from the $T=0$ ensembles with truncated
correlators as described in section~\ref{sec:quarkonium}.  Comparing
the two, we see clear evidence of a negative thermal mass shift for
$T\gtrsim140\,$MeV, well below the chiral transition temperature $T_c$.

We have also carried out a spectral reconstruction using the BR
method.  The results for the $B$ meson are shown in
fig.~\ref{fig:Bmesons} (right).  We find that the ground state peak
disappears around $T_c$, whereas the spectral function from the
equivalent truncated $T=0$ correlator still exhibits a clear peak,
suggesting that there is no bound state above $T_c$.

The analysis for the Gen2L ensembles is in progress, and preliminary
results support the conclusions presented above.

\section{Static quark potential}
\label{sec:SQP}

\begin{figure}[htb]
\includegraphics[width=0.49\textwidth,height=4.2cm]{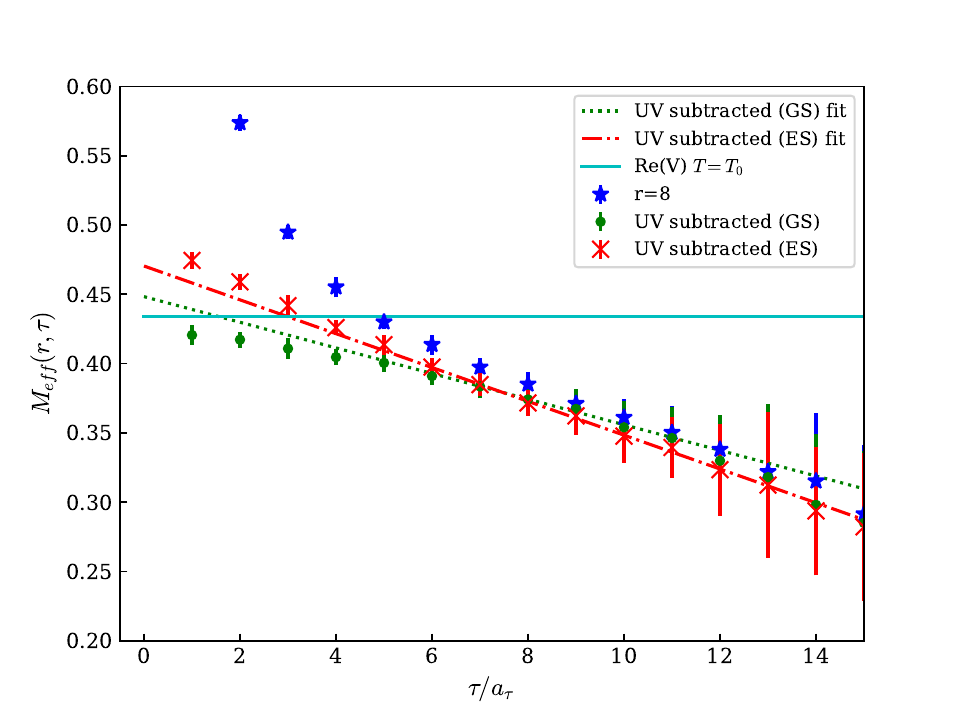}
\includegraphics[width=0.49\textwidth]{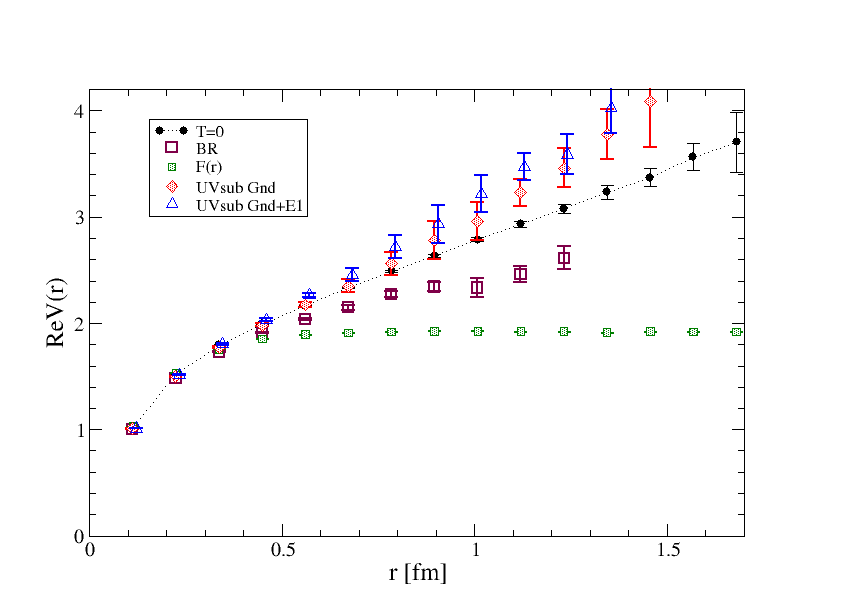}
\caption{Static quark potential from the Gen2L ensemble at $T/T_c=1.5
  (N_\tau=24)$ Left: Effective energies of the Wilson line correlator at
  $r/a_s=8$ from $T=0$, 1-state (GS) and 2-state (ES+GS) UV subracted
  correlators, together with fits to the UV subtracted correlators,
  assuming a single Gaussian form.  Right: Static quark potential from
the BR and UV-subtraction methods, compared with the $T=0$ potential
and the free energy.}
\label{fig:SQP}
\end{figure}

The static quark potential on the Gen2L ensemble has been computed
from Coulomb-gauge Wilson line correlators, which were computed with
the \textsc{simulat}e\textsc{qcd} code \cite{HotQCD:2023ghu}.  The
potential has been determined from the Wilson line correlators using
two methods: BR spectral function reconstruction, and the ``UV
subtraction'' method presented in \cite{Bazavov:2023dci}.  In the
latter method, the assumption is that the spectral function consists
of a ground state peak plus a temperature-independent ``UV'' part.
The UV contribution to the correlator can be easily isolated at zero
temperature and subtracted from the correlator at all temperatures.
The resulting ground state peak is then modelled by a gaussian, which
corresponds to an effective mass as a function of $\tau$ describing a
straight line, with the slope giving the imaginary part of the
potential and the intercept at $\tau=0$ giving the real part.

We have also modified this method by assuming that the spectrum may
contain a temperature-dependent first excited state in addition to the
ground state.  A sample of the results, using both assumptions, is
shown in fig.~\ref{fig:SQP} (left panel).  Here we see that at
$T\approx1.5T_c$ and $r\approx0.9\,$fm, the UV-subtracted effective
mass does not describe a straight line (invalidating the assumption of
a single gaussian), and a fit to a straight line
yields a value for $\Re V$ which is larger than the $T=0$ value.
Including the excited state yields a value which lies even higher.
The resulting values for $\Re V$ are shown in fig.~\ref{fig:SQP}
(right), together with the $T=0$ potential, the static
quark--antiquark free energy, and the results from the BR method.

Whereas the UV subtraction method applied to our data suggests
anti-screening at high temperature, the BR results indicate screening,
and are inconsistent with both the $T=0$ potential and the free
energy.  We are currently investigating modifications of the UV
subtraction method to better model the shape of the ground state peak,
which should be a skewed lorentzian rather than a gaussian.

\section*{Acknowledgments}
This work is supported by STFC grant ST/X000648/1. RHD acknowledges
support from a Maynooth University John and Pat Hume award and a
Research Ireland scholarship.  GA is supported by a Royal Society
Leverhulme Trust Senior Research Fellowship. RB acknowledges support
from a Science Foundation Ireland Frontiers for the Future Project
award with grant number SFI-21/FFP-P/10186. SK is supported by the
National Research Foundation of Korea through the grant,
NRF-2008-000458.  AR acknowledges support from Korea University via
project K2605081 ``Ab-initio lattice simulations of the real-time
dynamics of non-relativistic fermions''.  We acknowledge the use of
computing resources from the Irish Centre for High-End Computing
(ICHEC) and Supercomputing Wales.  This work used the DiRAC Data
Intensive service (DIaL2 \& DIaL) at the University of Leicester,
managed by the University of Leicester Research Computing Service on
behalf of the STFC DiRAC HPC Facility (www.dirac.ac.uk), the DiRAC
Extreme Scaling service (Tursa) at the University of Edinburgh,
managed by the Edinburgh Parallel Computing Centre on behalf of DiRAC,
and PRACE resources at CINECA (Marconi).

\bibliographystyle{JHEP}
\bibliography{eqcd-procs}

@article{Burnier:2013nla,
    author = "Burnier, Yannis and Rothkopf, Alexander",
    title = "Bayesian Approach to Spectral Function Reconstruction for Euclidean Quantum Field Theories",
    eprint = "1307.6106",
    archivePrefix = "arXiv",
    primaryClass = "hep-lat",
    doi = "10.1103/PhysRevLett.111.182003",
    journal = "Phys. Rev. Lett.",
    volume = "111",
    pages = "182003",
    year = "2013"
}

@inproceedings{Skullerud:2025iqt,
    author = "Skullerud, Jon-Ivar and others",
    title = "{Spectral properties of bottomonium at high temperature: a systematic investigation}",
    booktitle = "{16th Conference on Quark Confinement and the Hadron Spectrum}",
    eprint = "2503.17315",
    archivePrefix = "arXiv",
    primaryClass = "hep-lat",
    month = "3",
    year = "2025"
}

@article{Darcy:2025tzj,
    author = "D'Arcy, Rachel Horohan and others",
    title = "{NRQCD} Bottomonium at non-zero temperature using time-derivative moments",
    eprint = "2502.03951",
    archivePrefix = "arXiv",
    primaryClass = "hep-lat",
    doi = "10.22323/1.466.0203",
    journal = "PoS",
    volume = "LATTICE2024",
    pages = "203",
    year = "2025"
}

@article{Smecca:2025hfw,
    author = "Smecca, Antonio and others",
    title = "{The NRQCD $\Upsilon$ spectrum at non-zero temperatures using Backus-Gilbert regularisations}",
    eprint = "2502.03060",
    archivePrefix = "arXiv",
    primaryClass = "hep-lat",
    doi = "10.22323/1.466.0197",
    journal = "PoS",
    volume = "LATTICE2024",
    pages = "197",
    year = "2025"
}

@article{Bignell:2025bga,
    author = "Bignell, Ryan and others",
    title = "{Anisotropic excited bottomonia from a basis of smeared operators}",
    eprint = "2502.03185",
    archivePrefix = "arXiv",
    primaryClass = "hep-lat",
    doi = "10.22323/1.466.0202",
    journal = "PoS",
    volume = "LATTICE2024",
    pages = "202",
    year = "2025"
}

@article{Aarts:2014nba,
      author         = "Aarts, Gert and Allton, Chris and Amato, Alessandro and
                        Giudice, Pietro and Hands, Simon and Skullerud, Jon-Ivar",
      title          = "Electrical conductivity and charge diffusion in thermal
                        {QCD} from the lattice",
      journal        = "JHEP",
      volume         = "1502",
      pages          = "186",
      doi            = "10.1007/JHEP02(2015)186",
      year           = "2015",
      eprint         = "1412.6411",
      archivePrefix  = "arXiv",
      primaryClass   = "hep-lat",
      reportNumber   = "HIP-2014-34-TH, INT-PUB-14-060, MS-TP-14-40",
      SLACcitation   = "%%CITATION = ARXIV:1412.6411;%%",
}

@article{Aarts:2020vyb,
    author = "Aarts, G. and others",
    title = "Properties of the {QCD} thermal transition with {$N_f=2+1$} flavours of Wilson quark",
    eprint = "2007.04188",
    archivePrefix = "arXiv",
    primaryClass = "hep-lat",
    doi = "10.1103/PhysRevD.105.034504",
    journal = "Phys. Rev. D",
    volume = "105",
    number = "3",
    pages = "034504",
    year = "2022"
}

@article{Aarts:2022krz,
    author = {Aarts, Gert and Allton, Chris and Bignell, Ryan and Burns, Timothy J. and Garc\'\i{}a-Mascaraque, Sergio Chaves and Hands, Simon and J\"ager, Benjamin and Kim, Seyong and Ryan, Sin\'ead M. and Skullerud, Jon-Ivar},
    title = "{Open charm mesons at nonzero temperature: results in the hadronic phase from lattice QCD}",
    eprint = "2209.14681",
    archivePrefix = "arXiv",
    primaryClass = "hep-lat",
    month = "9",
    year = "2022"
}

@article{Aarts:2014cda,
      author         = "Aarts, Gert and Allton, Chris and Harris, Tim and Kim,
                        Seyong and Lombardo, Maria Paola and others",
      title          = "The bottomonium spectrum at finite temperature from
                        {$N_{f}=2+1$} lattice {QCD}",
      journal        = "JHEP",
      volume         = "1407",
      pages          = "097",
      doi            = "10.1007/JHEP07(2014)097",
      year           = "2014",
      eprint         = "1402.6210",
      archivePrefix  = "arXiv",
      primaryClass   = "hep-lat",
      SLACcitation   = "%%CITATION = ARXIV:1402.6210;%%",
}

@article{Edwards:2008ja,
      author         = "Edwards, Robert G. and Jo{\'o}, Balint and Lin, Huey-Wen",
      title          = "Tuning for Three-flavors of Anisotropic Clover Fermions
                        with Stout-link Smearing",
      journal        = "Phys.Rev.",
      volume         = "D78",
      pages          = "054501",
      doi            = "10.1103/PhysRevD.78.054501",
      year           = "2008",
      eprint         = "0803.3960",
      archivePrefix  = "arXiv",
      primaryClass   = "hep-lat",
      reportNumber   = "JLAB-THY-08-806",
      SLACcitation   = "%%CITATION = ARXIV:0803.3960;%%",
}

@article{Lin:2008pr,
      author         = "Lin, Huey-Wen and others",
      title          = "{First results from 2+1 dynamical quark flavors on an
                        anisotropic lattice: Light-hadron spectroscopy and setting
                        the strange-quark mass}",
      collaboration  = "Hadron Spectrum Collaboration",
      journal        = "Phys.Rev.",
      volume         = "D79",
      pages          = "034502",
      doi            = "10.1103/PhysRevD.79.034502",
      year           = "2009",
      eprint         = "0810.3588",
      archivePrefix  = "arXiv",
      primaryClass   = "hep-lat",
      reportNumber   = "JLAB-THY-08-896, TCDMATH-08-13",
      SLACcitation   = "%%CITATION = ARXIV:0810.3588;%%",
}

@article{Bazavov:2023dci,
    author = "Bazavov, Alexei and Hoying, Daniel and Larsen, Rasmus N. and Mukherjee, Swagato and Petreczky, Peter and Rothkopf, Alexander and Weber, Johannes Heinrich",
    collaboration = "HotQCD",
    title = "{Unscreened forces in the quark-gluon plasma?}",
    eprint = "2308.16587",
    archivePrefix = "arXiv",
    primaryClass = "hep-lat",
    doi = "10.1103/PhysRevD.109.074504",
    journal = "Phys. Rev. D",
    volume = "109",
    number = "7",
    pages = "074504",
    year = "2024"
}

@article{HotQCD:2023ghu,
    author = "Mazur, Lukas and others",
    collaboration = "HotQCD",
    title = "{SIMULATeQCD: A simple multi-GPU lattice code for QCD calculations}",
    eprint = "2306.01098",
    archivePrefix = "arXiv",
    primaryClass = "hep-lat",
    doi = "10.1016/j.cpc.2024.109164",
    journal = "Comput. Phys. Commun.",
    volume = "300",
    pages = "109164",
    year = "2024"
}

@article{Andronic:2024oxz,
    author = "Andronic, A. and others",
    title = "{Comparative study of quarkonium transport in hot QCD
matter}",
    eprint = "2402.04366",
    archivePrefix = "arXiv",
    primaryClass = "nucl-th",
    reportNumber = "FERMILAB-PUB-24-0005-T-V",
    doi = "10.1140/epja/s10050-024-01306-6",
    journal = "Eur. Phys. J. A",
    volume = "60",
    number = "4",
    pages = "88",
    year = "2024"
}

\end{document}